C:/comets/Disintegrating Comets/3 Predictions/140306

**Submitted for publication**

# "Three Predictions:

# Comet 67P/Churyumov-Gerasimenko,

# Comet C/2012 K1 Panstarrs, and

# Comet C/2013 V5 Oukaimeden"

Ignacio Ferrín,
Institute of Physics,
Faculty of Exact and Natural Sciences,
University of Antioquia,
Medellin, Colombia, 05001000
ferrin@fisica.udea.edu.co

Number of pages     7

Number of Figures   4

Number of Tables    1

**Abstract**

We make the following predictions:

**(1) Comet 67P/Churyumov-Gerasimenko.** The Secular Light Curve (SLC) of this comet exhibits a photometric anomaly in magnitude that is present in 1982, 1996, 2002 and 2009. Thus it must be real. We interpret this anomaly as a topographic feature on the surface of the nucleus that may be a field of debris, a region made only of dust or an area of solid stones but in any case it is depleted in volatiles. *We predict that images taken by spacecraft Rosseta will show a region morphologically different to the rest of the nucleus, at the pole pointing to the Sun near perihelion.*

**(2) Comet C/2012 K1 Panstarrs.** This comet exhibits the same Slope Discontinuity Event (SDE)+magnitude dip after the event than 11 other comets listed in Table 1 all of which disintegrated. This group includes comet C/2012 S1 ISON. Thus it is reasonable to expect that this comet may disintegrate too. *The probability of disintegration of this comet is 20%.*

**(3) Comet C/2013 V5 Oukaimeden.** This comet exhibits the same SDE+dip signature exhibited by 11 other comets in Table 1 all of which disintegrated. This group includes comet C/2012 S1 ISON. *We predict that there is a 90% probability that this comet will disintegrate.*

Key words: comets; solar system

Highlights:

1) We predict that Comet 67P/Churyumov-Gerasimenko will exhibit a topographic feature at one of the poles, in the upcoming images taken by the Rosseta spacecraft in August of this year.
2) We predict that comet C/2012 K1 Panstarrs may disintegrate, and the probability of this happening is 20%.
3) We predict that comet C/2013 V5 Oukaimeden may also disintegrate with a probability of 90%.

**1. Introduction**

An important part of the Scientific Method establishes that predictive power is a required component of any scientific theory. Predictions test our capability to understand the Universe. Recently the research line on Secular Light Curves of Comets (SLCs) has been advanced significantly (Ferrín, 2005, 2006, 2007, 2008, 2009, 2010, 2013a, 2013b, 2013c). The SLCs are the scientific way to study the brightness behavior of a comet, and more than 20 physical parameters, most of them new, can be extracted from the light curves. SLCs allow making predictions on comets as we will see.

In the *Atlas of Secular Light Curves of Comets*, V.1, (Ferrín, 2010), several light curves of comets showed photometric anomalies that do not have a physical explanation. One of those comets was 67P/Churyumov-Gerasimenko. In this work we interpreted the photometric anomaly of this comet as a topographic feature on the surface of the nucleus.

Advances in our SLCs comprehension have allowed to predict the disintegration of comet C/2012 S1 ISON (Ferrín 2013a, 2013b, 2013c). The prediction was based on the fact that the comet exhibited a Slope Discontinuity Event (SDE) + a magnitude-dip-after-the-event signature that appears also in 11 other comets listed in Table 1, all of which disintegrated. Thus we will make two predictions on the disintegration of comets C/2012 K1 Panstarrs, and Comet C/2013 V5 Oukaimeden, and will give a probability for their occurrence.

**2. Comet 67P/Churyumov-Gerasimenko**

There has been a recent determination of the diameter of this comet by Kelley et al. (2009) and it is adopted in the SLC presented in Figure 1. The SLC of this comet exhibits a dip in magnitude that is reproduced in 1982, 1996, 2002 and 2009. Thus it must be real. There has not been a physical interpretation of this feature and we are now proposing that a possible explanation is a topographic feature on the surface of the nucleus. The feature may be a field of debris, a region made only of dust or an area of solid stones but in any case it has to be an area depleted in volatiles. Spacecraft mission Rosseta is due to reach the comet in middle 2014 and will take high resolution images of the nucleus. *Thus our prediction is that these images will show a region of different morphological characteristics to the rest of the nucleus at the pole pointing to the Sun near perihelion.*

If the north or south pole of the comet pointed to the Sun at ~ -68 days before perihelion (obliquity of 90º), then there would be a polar cap of ~54º in radius, depleted in volatiles by a factor of 5.3x. On the other hand if the pole has an obliquity of 45º, then the depleted area may extend to half of the nucleus, producing two unequal hemispheres with different terrains. If the equator of the comet pointed all the time to the Sun (zero obliquity), there would be no way to explain the observations. Thus the photometric observations suggest that the comet has a large obliquity.

This prediction will be tested when spacecraft Rosseta reaches to the comet and takes high resolution images of its surface (http://sci.esa.int/rosetta/2279-summary/ ). The spacecraft will arrive at the comet on August 2014 when it will start a global mapping, and a lander will be deployed in November of the same year.

**3. Comet C/2012 K1 Panstarrs**

This comet exhibits the same Slope Discontinuity Event (SDE)+ magnitude dip after the event than 14 other comets listed in Table 1 that disintegrated, including the famous C/2012 S1 ISON that was predicted to disintegrate (Ferrín 2013a, 2013b, 2013c) and which in fact disintegrated (CBET 3731). Thus it is reasonable to expect that this comet may disintegrate too. However, the perihelion distance of this comet is q = 1.05 AU, and thus it does not reach to the critical distance R(critical) = 0.63 AU deduced from Table 1. *From the perihelion and disintegration distances listed in Table 1 it is possible to predict that the probability of disintegration of this comet is only 20%.*

This prediction will be tested around August 27$^{th}$, 2014, when the comet will reach perihelion. However the comet will be immersed in the solar glare. From August 6$^{th}$ to August 14$^{th}$, the elongation angle from the Sun will be E < 4º which implies that it should appear in images taken by spacecrafts observing the Sun, like SOHO and Stereo. On August 27$^{th}$, the perihelion date, the elongation angle will be E = 18º, so it may be observed for a brief period of time from ground based observatories.

**4. Comet C/2013 V5 Oukaimeden**

This comet exhibits the same SDE+dip signature exhibited by 14 other comets in Table 1 that disintegrated, including the famous comet C/2012 S1 ISON that was predicted to disintegrate (Ferrín 2013a, 2013b, 2013c) and which in fact disintegrated (CBET 3731). This is a dynamically new comet (DNC) and from Table 1 it is clear that DNCs have a large probability of disintegrating. This result plus a perihelion distance q = 0.62 AU < R(critical) = 0.63 AU deduced from Table 1, makes it virtually certain that this comet is going to disintegrate. The most probable distance of disintegration is near perihelion. Unfortunately the observational circumstances will be complicated by the fact that the comet remains most of the apparition below an elongation angle E < 40º. *For this comet our prediction is that the probability of disintegration is 90%.*

This prediction will be tested around September 27$^{th}$, 2014, when the comet will reach perihelion. On that date the elongation angle will be E = 38º, thus the comet will be observable for a brief period of time from ground based telescopes.

Due to space limitations it is impossible in this note to go into theoretical details of why the SDE+dip signature is a good predictor of disintegration. This subject will be studied in a different paper.

**5. Conclusions**

Scientific understanding is tested by its predictions. In this work we make three predictions using the powerful methodology provided by the Secular Light Curves of Comets. Comet 67P/Churyumov-Gerasimenko has a photometric anomaly that could be caused by a topographic anomaly on the surface of the comet, specifically at one of the poles. This prediction will be tested when the spacecraft Rosseta reaches to the comet and takes high resolution images of its surface, on August of 2014.

For comets C/2012 K1 Panstarrs and C/2013 V5 Oukaimeden, we predict a high probability of disintegration of 20% and 90%, respectively. The disintegration may follow the same pattern exhibited by comet C/2012 S1 ISON.

It is becoming clear that dynamically new comets are fragile objects that have a tendency to disintegrate (confirmation Table 1). The disintegration probability depends on their diameter and its perihelion distance. However up to know we have SLCs of 14 objects and all have been sufficiently small and near to the Sun to disintegrate.

Updates on these objects will be placed at the web site:
http://astronomia.udea.edu.co/cometspage

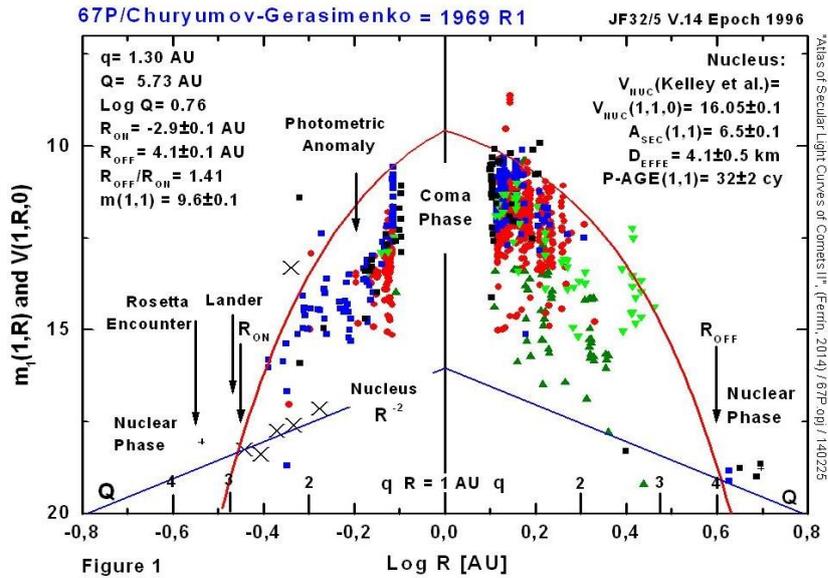

*Figure 1. The Secular Light Curve of Comet 67P/Churyumov-Gerasimenko.* *The SLC exhibits a photometric anomaly that is interpreted as a topographic feature on the surface of the nucleus. The feature repeats in the same place and at the same time from perihelion in four apparitions, 1982, 1996, 2002 and 2009. The recent determination of the diameter of this comet, D = 4.08 km is adopted in this plot (Kelley et al. 2009). The data for this plot and the next comes from Green (2013).*

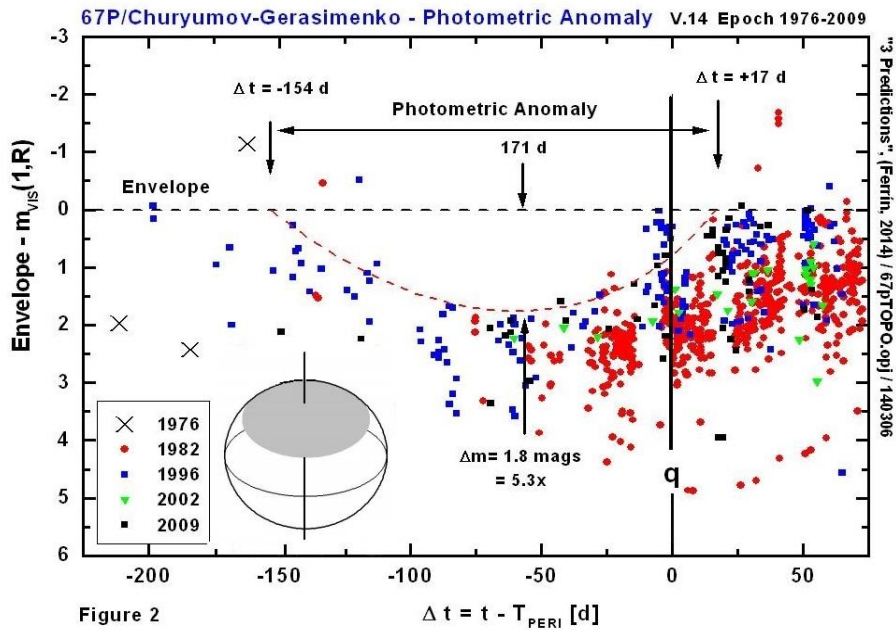

*Figure 2. The envelope of the data set has been subtracted from the observations and now the light curve is flat.* *The photometric anomaly is clearly seen and it is reproduced in four apparitions, 1982, 1996, 2002 and 2009. The feature can be explained if the pole of the comet points to the Sun at ~ -68 days before perihelion and if it is depleted in volatiles. The true anomaly is $v = -93.9°$ at $\Delta t = -154$ d and $v = +15.0°$ at $\Delta t = +17$ d. In the inset the shape and location of the region is displayed.*

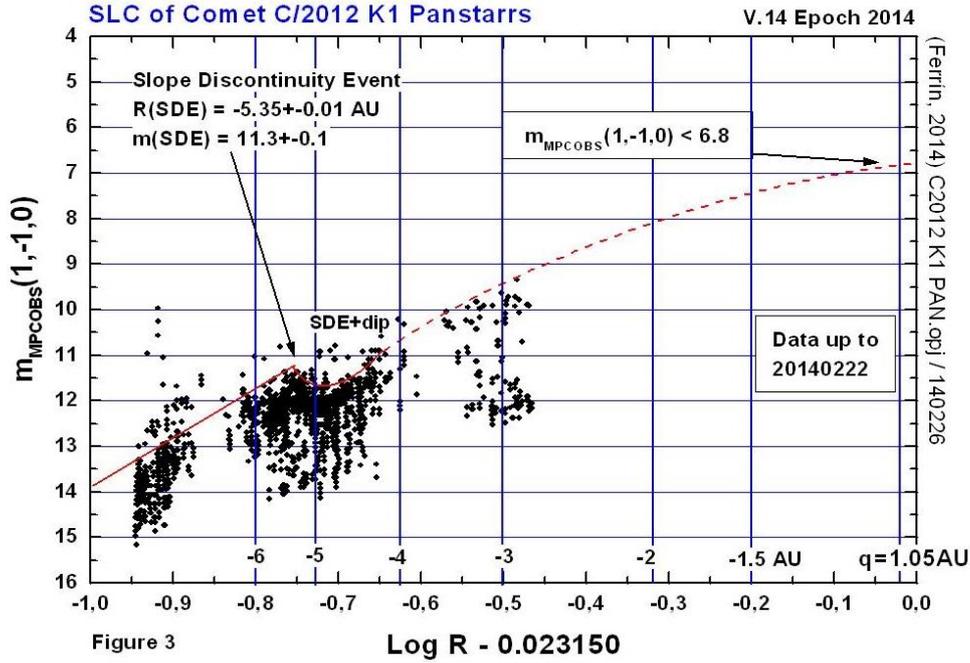

*Figure 3. The Secular Light Curve of Comet C/2012 K1 Panstarrs. It exhibits the Slope Discontinuity Event+Dip signature of comets that have disintegrated in the past (Table 1), and thus there is a probability that this comet will also disintegrate. However since the perihelion distance of this comet is q = 1.05 AU the probability of disintegration is only 20%. The data for this plot comes from the Minor Planet Center Dataset (Spahr 2013).*

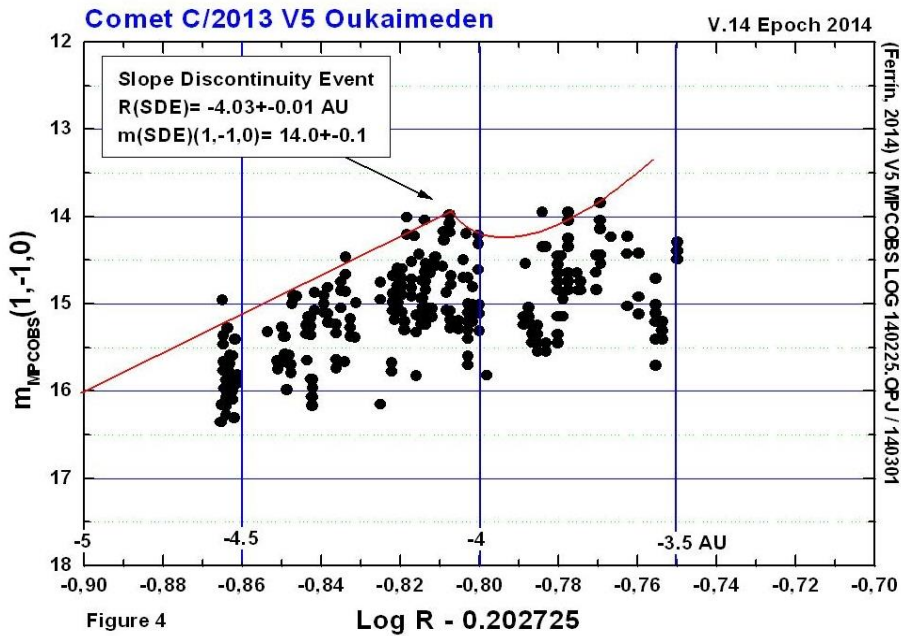

*Figure 4. The Secular Light Curve of Comet C/2013 V5 Oukaimeden. The comet exhibits the same SDE+Dip signature of other comets that have disintegrated (Table 1). Additionally the comet is dynamically new. The probability that this comet will disintegrated is 100%. During 2014 the object will expend most of its time with elongation angles E < 40º. Thus confirming its disintegration will be challenging. The data for this plot comes from the Minor Planet Center Dataset (Spahr 2013).*

1. ISON's Group. (7)(8)(9)

```
--------------------------------------------------------------------------------
#  Comet                             R(DISIN)   q      CODE       1/a ORIGINAL Identifier
                                     [AU]       [AU]   (0)        or e (10)     (1)
--------------------------------------------------------------------------------
01 C/2002 O4 (Hönig)            (5)  -0.88      0.78   S+d+DN+D   -0.00077249  I. Ferrín
02 C/2012 T5 (Bressi)                -0.66      0.32   S+d+DN+D   -0.00022724  J. Cerny
03 C/1999 S4 (LINEAR)                -0.92      0.77   S+d+DN+D   -0.00005451  G. Kronk
04 C/2013 V5 (Oukaimeden)       (6)  -----      0.63   S+d+DN     -0.00003227  I. Ferrín
05 C/2009 R1 (McNaught)         (4)  -0.79      0.41   S+d+DN+D   -0.00003180  C. Hergenrother
06 C/2012 V1 (Panstarrs)             -----      2.09   S+d+DN+D   -0.00001928  I. Ferrín
07 C/2012 S1 (ISON)            (11)  -0.67      0.01   S+d+DN+D    0.00000852  I. Ferrín
08 C/2012 K1 (Panstarrs)        (6)  -----      1.05   S+d         0.00001807  I. Ferrín
09 C/2010 X1 (Elenin)                -0.64      0.48   S+d+D       0.00011071  A. Cook
10 C/1996 Q1 (Tabur)                 -0.93      0.84   S+d+D       0.00182608  I. Ferrín
11 C/1897 U1 (Perrine)= 1897 III     -1.61      1.36   D           1.0000000   Z. Sekanina
12 C/1957 U1 (Latyshev-Wild-Burnham) -1.11      0.54   D           1.0000000   Z. Sekanina
13 C/1974 V2 (Bennet) 1974XV         -0.98      0.86   D           1.0000000   Z. Sekanina
14 C/1953 X1 (Pajdusakova) 1954 II   -0.74      0.07   D           1.000000    Z. Sekanina
15 C/2008 J4 (McNaught) Headless comet -0.64    0.45   D           1.000000    I. Ferrín
16 C/1925 X1 (Ensor) = 1926 III      -0.63      0.32   D           1.000000    Z. Sekanina
17 C/1887 B1 headless comet     (2)  -----      0.005  D           1.0000000   Z. Sekanina
18 20D/1913 S1 (Westphal)            -1.43      1.25   D           0.9198301   Z. Sekanina
19 P/2006 HR30 (Siding Spring)  (3)  -----      1.23   S+d         0.8437940   I. Ferrín
20 C/1997 N1 (Tabur)(unconfirmed)    -----      0.40   D           1.0001344   G. Kronk
--------------------------------------------------------------------------------
```

0. CODE: S=Slope Discontinuity Event (SDE); d=dip after SDE; DN = Dynamically New; D = Disintegrated. Additionally the group has -1.61 < R(DIS) < -0.63 AU.
1. After publishing the first manuscript in the Arxiv.org depository, Gary Kronk, Toni Cook and Jacub Cerny (personal communication), discovered four additional members of the group.
2. See Sekanina (1984).
3. All comets in this list are Oort Cloud members, while this comet is periodic, which raises some interesting questions about its origin. Additionally this comet has not disintegrated and it is due to return in 2027. However it exhibits the signature.
4. Hergenrother (2010) has a light curve showing the SDE+dip, while Seiichi's web site http://www.aerith.net shows that the comet was not detected post-perihelion, all consistent with disintegration, although this was not observed.
5. Dynamically new comets have been highlighted in black.
6. These two comets exhibit the SDE+dip signature, but are still far from perihelion and have not yet disintegrated.
7. From Column 3, R(Disintegration), we deduce that there is a R(limit)= -0.63 AU (Pre-perihelion). All comets have disintegrated before reaching this limit.
8. From the next to the last column, it is clear that dynamically new comets have a clear tendency to disintegrate.
9. The SLCs of many of these comets have been presented by Ferrín (2013a, 2013b, 2013c) and are not repeated here due to space limitations.
10. In this work we define dynamically new as 1/a(original) < 0.00001000. The 1/a(original) value is given when available.
11. The disintegration distance for comet ISON was determined using data by Ferrín (2013a) and by Combi et al. (2013) (CBET 9266).